\documentclass[prl,twocolumn,showpacs,floatfix]{revtex4}
\usepackage{graphicx}

\begin{document}
\title{The Scaling Behavior of Classical Wave Transport in Mesoscopic Media at the Localization Transition}

\author{S.K.~Cheung and Z.Q.~Zhang}

\address{Department of Physics, Hong Kong University of Science and Technology, Clear Water Bay, Kowloon, Hong Kong}

\date{\today}

\begin{abstract}
The propagation of classical wave in disordered media at the Anderson localization transition is studied. Our results show that the
classical waves may follow a different scaling behavior from that for electrons. For electrons, the effect of weak localization due to
interference of recurrent scattering paths is limited within a spherical volume because of electron-electron or electron-phonon scattering,
while for classical waves, it is the sample geometry that determine the amount of recurrent scattering paths that contribute. It is found
that the weak localization effect is weaker in both cubic and slab geometry than in spherical geometry. As a result, the averaged static
diffusion constant $D(L)$ scales like $\ln(L)/L$ in cubic or slab geometry and the corresponding transmission follows $\langle
T(L)\rangle\propto\ln L/L^2$. This is in contrast to the behavior of $D(L)\propto 1/L$ and $\langle T(L)\rangle\propto 1/L^2$ obtained
previously for electrons or spherical samples. For wave dynamics, we solve the Bethe-Salpeter equation in a disordered slab with the
recurrent scattering incorporated in a self-consistent manner. All of the static and dynamic transport quantities studied are found to
follow the scaling behavior of $D(L)$. We have also considered position-dependent weak localization effects by using a plausible form of
position-dependent diffusion constant $D(z)$. The same scaling behavior is found, i.e., $\langle T(L)\rangle \propto\ln L/L^2$.
\end{abstract}

%42.25.Dd - Wave propagation in random media
%42.25.Bs - Wave propagation, transmission, and absorption
%72.15.Rn - Localization effects (Anderson or weak localization)
%72.20.Ee - Mobility edges; hopping transport

\maketitle
\begin{section}{I. Introduction}
The Anderson localization transition in three dimensions occurs when the Ioffe-Regel criterion $k\ell_0\lesssim1$ is met, where $k$ is the
wave vector and $\ell_0$ is the bare mean free path \cite{Sheng}. In the diffusive regime, it is known that the averaged transmission
coefficient decays with the sample size like $\langle T(L)\rangle \simeq\ \ell_0/L$.   In the localized regime, it is also known that the
geometrical mean of the transmission coefficient falls off exponentially with $L$, i.e. $\langle T(L)\rangle _{g} \simeq\exp(-L/\xi)$, where
$\xi$ is the localization length. At the localization transition, it has been predicted that $\langle T(L)\rangle\simeq(\ell_0/L)^2$
\cite{Anderson85}. This $1/L^2$ behavior is obtained by considering the reduction of the Boltzmann diffusion constant, $D_0$, due to weak
localization (WL) effects in a spherical volume of size $L^3$.  It arises from the contributions of all recurrent scattering paths returning
to the origin inside the volume, i.e., those paths of length longer than $L$ do not contribute to the reduction of $D_0$.  As a result, the
renormalized diffusion constant, $D(L)$, becomes size-dependent and has the form $D(L)\simeq D_0\ell_0/L$ at the localization transition.
Equivalently, the mean free path is renormalized to $\ell \simeq \ell_0^2/L$. This renormalization of intrinsic transport parameters gives
rise to the scaling behavior of $\langle T(L)\rangle \propto D(L)/L \propto 1/L^2$ \cite{Anderson85, Genack90}. In one and two dimensions,
the recurrent scattering paths give divergent contributions to the reduction of $D_0$ when $L$ is large, i.e., $\delta D \propto L$ in 1D
and $\delta D \propto \ln L$ in 2D \cite{Sheng}. Thus all states are believed to be localized in one- and two-dimensional random media
\cite{Abrahams79}. For electrons, inelastic scattering due to electron-electron or electron-phonon interaction provides a nature cutoff
length for the recurrent scattering paths. Since the inelastic scattering time is inversely proportional to some power in the temperature,
i.e., $\tau_{in} \propto T^{-p}$, WL leads to a temperature-dependent conductivity which decreases like $-T^{-p/2}$ in 1D and $p\ln T$ in 2D
as long as the dephasing length $L_{dep}\propto\sqrt{\tau_{in}}$ is smaller than the sample length $L$. Such temperature-dependent of
conductivity has been observed in disordered metal wires and films \cite{Sheng,Bergmann}. For classical waves, the observation of light
localization and the scaling behavior of $\langle T(L)\rangle\simeq(\ell_0/L)^2$ have been reported \cite{Wiersma97,Garcia91}. However,
these reports have come under close scrutiny because of the presence of absorption in the samples.

Here we would like to point out that the renormalization of $D_0$ discussed above may not be valid for classical waves or for electrons when
the dephasing length $L_{dep}$ is larger than the sample size $L$. In this case, the sample geometry determines the cutoff of the recurrent
scattering paths. For classical waves, samples used for the transmission measurements are usually not in spherical geometry
\cite{Wiersma97,Garcia91}. For example, slab geometry are often adapted in optical \cite{Wiersma97} and ultrasonic measurements
\cite{Zhang99}. Since the recurrent scattering paths in a slab are different from those in a spherical sample, it is natural to ask whether
the previously obtained scaling behaviors for electrons are actually applicable to classical waves? If not, what should be correct scaling
behaviors for classical waves? The purpose of this work is to address these questions.

In this work, we study the propagation of classical waves in finite size disordered samples at the localization transition or mobility edge
under the framework of the self-consistent theory of localization \cite{Vollhardt82,Kirkpatrick85}. The contributions from all recurrent
scattering paths within the slabs are calculated in the framework of self-consistent theory of localization. We show that the averaged
static diffusion constant $D(L)$ is proportional to $\ln L/L$ at the mobility edge for both cubic, cylindrical and slab geometries, in
contrast to to the behavior of $D(L)\propto 1/L$ obtained previously for elecrons or spherical samples \cite{Anderson85}.  The corresponding
static transmission follows the scaling $\langle T(L)\rangle\propto\ln L/L^2$.  For dynamics, we have studied the time-dependent wave
propagation in disordered slabs by using both the Bethe-Salpeter equation and the diffusion equation with a frequency-dependent diffusion
constant. In both equations, the effects due to WL are incorporated in a way that renormalizes the mean free path \cite{Zhang04}. It will be
shown that the diffusion equation produces the same scaling behavior as that of the B-S equations when $L\gg\ell_0$.

We have also consider the position-dependent WL effects by using a plausible form of position-dependent diffusion constant $D(z)$. We find
that the scaling behavior $\langle T(L)\rangle\propto\ln L/L^2$ holds when $L/\ell_0$ is large. Since the localization effect studied in
this work is a general wave phenomena, these new scaling laws are not limited to classical waves, but may also apply to electrons if
$L_{dep}$ is larger than the sample size $L$.
\end{section}

\begin{section}{II. Theory}
\begin{subsection}{A. Scaling behavior of renormalized averaged diffusion constant at mobility edge}
Weak localization due to interference of recurrent scattering paths can be signified by the reduction of diffusion constant in the frequency
domain. By summing all the maximally-crossed diagrams \cite{Lagendijk88} in self-consistent diagrammatic theory, the renormalized diffusion
constant in a bulk can be written as \cite{Vollhardt82,Kirkpatrick85}

\begin{equation}
\frac{1}{D(\omega,k)}={\frac{1}{D_0}} [1+\frac{2\pi v}{k^2}\tilde{G}(\omega;\mathbf{r},\mathbf{r})], \label{}
\end{equation}
where $D(\omega,k)=v\ell(\omega,k)/3$, $D_0=v\ell_0/3$ is the Boltzmann diffusion coefficient and $\tilde{G}$ is the Green's function that
satisfies the diffusion equation in the frequency domain:
\begin{equation}
(D_0\nabla^2+i\omega)\tilde{G}(\omega;\mathbf{r},\mathbf{r}')=-\delta(\mathbf{r}-\mathbf{r}').
\end{equation}
The diagonal term of the Green's function, $\tilde{G}(\omega;\mathbf{r},\mathbf{r}')$, represents the return probability of waves that
travel diffusively in the bulk. For an infinite medium, $D(\omega)$ has already been studied previously in different dimensions
\cite{Gorkov,Gotze,Shapiro,Vollhardt80,Weaver05}. To study the scaling behavior in a slab, we first solve for $\tilde{G}$ in a cylinder of
length $L$ and radius $R\gg\ell_0$ with open ends in cylindrical geometry. A slab can then be obtained by taking the limit of $R/L
\rightarrow \infty$. Later we will explain that either the cylindrical geometry or the cubic geometry would give the same scaling behavior
that is different from that in spherical geometry. The result can be written as

\begin{eqnarray}
\tilde{G}(\omega;z=z')=\frac{1}{2 \pi^2 \tilde{L}}\sum_{n=1}^{n_c}\sin^2[q_n(z+z_e)] \nonumber\\
\times\int_{\frac{1}{R}}^{\frac{\alpha}{\ell_0}}\frac{2\pi q_{\parallel}d q_{\parallel}}{-i\omega+D_0 (q_n^2+q_\parallel^2)},
\end{eqnarray}
where $q_n=n\pi/\tilde{L}$, $n_c=\alpha\tilde{L}/\pi l_0$ is the upper momentum cutoff in the z-direction, $\mathbf{q_{\parallel}}$ is the
momentum in the x-y plane, $\tilde{L}=L+2z_e$ is the effective thickness of the slab and $z_e\simeq0.7104\ell_{0}$ is the extrapolation
length \cite{Sheng}. We let $\alpha=1$ in our calculations. A different choice of $\alpha$ will only change the mobility edge, $k_c$, not
the scaling behaviors of wave transport. Eq. (3) indicates that the renormalized diffusion constant is $z$-dependent. In order to simplify
our calculations, we take the spatial average along the $z$-axis and replace the factor of $\sin^{2}[q_{n}(z+z_{0})]$ by $1/2$. This
averaging replaces the position dependent diffusion constant by its harmonic mean. The situation of $z$-dependent diffusion constant will be
considered in Sec. IIC.

Here we define a renormalization factor $\delta_L(\omega,k)$ that renormalize the diffusion constant in a finite size slab of thickness $L$
according to

\begin{equation}
D_L(\omega,k)=\frac{v\ell_L(\omega,k)}{3}=\frac{D_0}{1+\delta_L(\omega,k)},
\end{equation}
where $\delta_L(\omega,k)\equiv2\pi v \tilde{G}/k^2$ by comparing Eq. (4) with Eq. (1), and $\ell_L(\omega,k)=\ell_0/[1+\delta_L(\omega,k)]$
is the renormalized mean free path. By using Eqs. (3) and (4), we obtain

\begin{equation}
\delta_L(\omega,k) = \frac{v}{2k^2 \tilde{L}}\sum^{n_c}_{n=1}\int_{\frac{1}{R}}^{\frac{1}{\ell_0}}\frac{2q_{\parallel} d
q_{\parallel}}{-i\omega+D_0 (q_n^2+q_\parallel^2)}.
\end{equation}
Eq. (5) is then solved self-consistently by replacing $D_0$ with $D_L(\omega,k)$ \cite{Vollhardt82,Kirkpatrick85}. Physically, the
self-consistency of $\ell_L(\omega,k)$ or $D_L(\omega,k)$ assures the successive renormalization of the recurrent scattering paths inside
the samples.

The scaling properties of disordered slabs can be obtained by investigating the the scaling behavior of the renormalization factor
$\delta_L(0,k)$ in the static limit. When $L\gg\ell_0$, the summation in Eq. (5) can be replaced by an integral, leading to

\begin{equation}
\delta_L(0,k)\simeq\frac{v}{2\pi k^2
D_L(k)}\int_{L^{-1}}^{\ell_0^{-1}}dq_{\perp}\int_{R^{-1}}^{\ell_0^{-1}}\frac{2q_{\parallel}dq_\parallel}{q_\perp^2+q_\parallel^2}
\end{equation}
where $D_L(k)\equiv D_L(0,k)$ and $dq_\perp\equiv \Delta q_n$. Each of the double integrals in Eq. (6) can be split into two parts, i.e.

\begin{eqnarray}
\int_{L^{-1}}^{\ell_0^{-1}}\int_{R^{-1}}^{\ell_0^{-1}}=\left(\int_0^{\ell_0^{-1}}-\int_0^{L^{-1}}\right)\cdot\left(\int_0^{\ell_0^{-1}}-\int_0^{R^{-1}}\right),\nonumber
\end{eqnarray}
and thus Eq. (6) can be expressed as the sum of four terms:

\begin{equation}
\delta_L(0,k)\simeq\frac{v}{2\pi k^2 D_L(k)}(\eta_A+\eta_B+\eta_C+\eta_D),
\end{equation}
where

\begin{eqnarray}
\eta_A&\simeq&(\ln2+\frac{\pi}{2})/\ell_0\nonumber\\
\eta_B&\simeq&-(2\ell_0/L)[\ln\left(L/\ell_0\right)+1]\nonumber\\
\eta_C&\simeq&-(\ell_0/R^2+\pi/R)\\
\eta_D&\simeq& L/R^2+\pi/R\nonumber.
\end{eqnarray}
The mobility edge in a bulk can be obtained by taking the limits of $L\rightarrow\infty$ and $R\rightarrow\infty$ in Eq. (8). At such limit,
$\eta_B=\eta_C=\eta_D\equiv0$ and Eq. (7) becomes

\begin{equation}
\delta_{\infty}(0,k)= \frac{v}{2\pi k^2 D_\infty(k)\ell_0}\left(\ln2+\frac{\pi}{2}\right),
\end{equation}
where $D_\infty(k)\equiv D_\infty(0,k)$. By substituting Eq. (9) into Eq. (4), we obtain the following expression for $D_\infty(k)$:

\begin{equation}
\frac{D_\infty(k)}{D_0}=1-\frac{3}{2\pi(k\ell_0)^2}\left(\ln2+\frac{\pi}{2}\right).
\end{equation}
For the convenience of discussions, here we set $\ell_0$ as the units of length and let $k$ to vary.  Since the Anderson transition occurs
when $D_\infty(k_c)=0$, Eq. (10) gives the mobility edge $k_c\ell_0\equiv\sqrt{\frac{3(\ln2+\pi/2)}{2\pi}}\simeq1.039$.  We can now rewrite
Eq. (10) as

\begin{equation}
\frac{D(k)}{D_0}=1-\left(\frac{k_c}{k}\right)^2.
\end{equation}
For either a 'cubic-like' sample ($R=L$) or a slab ($R\rightarrow\infty$), $\eta_B$ dominates and Eq. (7) can be approximately written as

\begin{equation}
\delta_L(0,k)\simeq\frac{D_0}{D_L(k)}\left(\frac{k_c}{k}\right)^2\left[1-\left(\frac{\ell_0}{L}\right)\frac{2\ln(L/\ell_0)}{\ln2+\pi/2}\right].
\end{equation}
By using the relation $D_0/D_L(k) = 1+\delta_L(0,k)$ in Eq. (12), it is easy to see that $\delta_L(0,k)\propto L/\ln L$ when $k=k_c$. By
substituting Eq. (12) into Eq. (4), the static diffusion constant $D_L(k)$ in a finite slab but with $L\gg\ell_0$ can be expressed as

\begin{equation}
\frac{D_L(k)}{D_0}\simeq 1-\left(\frac{k_c}{k}\right)^2\left[1-\left(\frac{\ell_0}{L}\right)\frac{2\ln(L/\ell_0)}{\ln2+\pi/2}\right].
\end{equation}
Eq. (13) gives  $D(L)\equiv D_L(k_c) \propto \ln L/L$ at $k=k_c$, which can also be obtained from Eq. (4) by using $\delta_L(0,k_c) \propto
L/\ln(L)$. In the following section, we will see that the scaling of $D(L)$ dictates the scaling behaviors of many measured static and
dynamic transport quantities. When $k\gtrsim k_c$, Eq. (13) gives the following $L$-dependent static diffusion constant for slabs:

\begin{equation}
\frac{D_L(k)}{D_0}\simeq \left\{\begin{array}{r@{\quad, \quad}l} (\ell_0/L)\ln(L/\ell_0) & L<\xi_s \\
(\ell_0/\xi_s)\ln(\xi_s/\ell_0) & L>\xi_s \end{array}\right.,
\end{equation}
where $\xi_s$ is the saturation thickness beyond which $D_L(k)$ becomes virtually independent of $L$. $\xi_s$ can be estimated by requiring
the $L$-independent term equal to the $L$-dependent term in Eq. (13), yielding

\begin{equation}
\frac{\ell_0}{\xi_s}\ln\left(\frac{\xi_s}{\ell_0}\right)\sim \frac{0.72(k-k_c)}{k_c}.
\end{equation}
Eq. (15) gives a scaling behavior of $\xi_s^{-1}\propto|k-k_c|\ln|k-k_c|$. The above results are different from those obtained previously
for electrons or spherical samples. For a spherical sample of radius $L$, Eq. (2) gives \cite{Kirkpatrick85, Imry80}

\begin{equation}
\delta_L(0,k)\simeq\frac{v}{4\pi^2 k^2 D_L(k)}\int_{L^{-1}}^{\ell_0^{-1}}\frac{d \mathbf{q}}{q^2}.
\end{equation}
By substituting Eq. (16) in Eq. (4), we obtain

\begin{equation}
\frac{D_L(k)}{D_0}\simeq1-\frac{3}{\pi}\frac{1}{(k\ell_0)^2}\left(1-\frac{\ell_0}{L}\right),
\end{equation}
from which we obtain $D(L)\equiv D_L(k_c)\propto\ell_0/L$ at $k=k_c$, and when $k\gtrsim k_c$,

\begin{equation}
\frac{D_L(k)}{D_0}\simeq \left\{\begin{array}{r@{\quad, \quad}l} \ell_0/L & L<\xi \\ \ell_0/\xi & L>\xi
\end{array}\right.,
\end{equation}
where $\xi$ is the correlation length and is proportional to $|k-k_c|^{-1}$ \cite{Bart99}.

\begin{figure}
\includegraphics [width=\columnwidth] {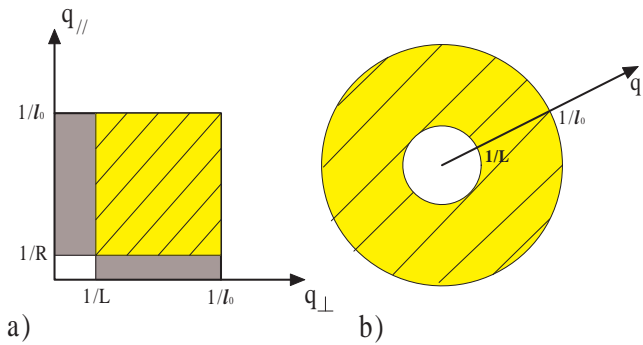}
\caption{The schematic diagrams to show the momentum spaces of diffusion for the (a) cylindrical and (b) spherical geometry. The shaded
areas represent the regions of momentum space that allow diffusion. For (a), the momentum space is a cylindrical volume and the rectangles
shown are the surfaces for volume of evolution about the $q_{\perp}$ axis. For (b), the momentum space is a spherical volume and the circles
shown are the cross-sections of the concentric spheres about the origin.}
\end{figure}

\begin{figure}
\includegraphics [width=\columnwidth] {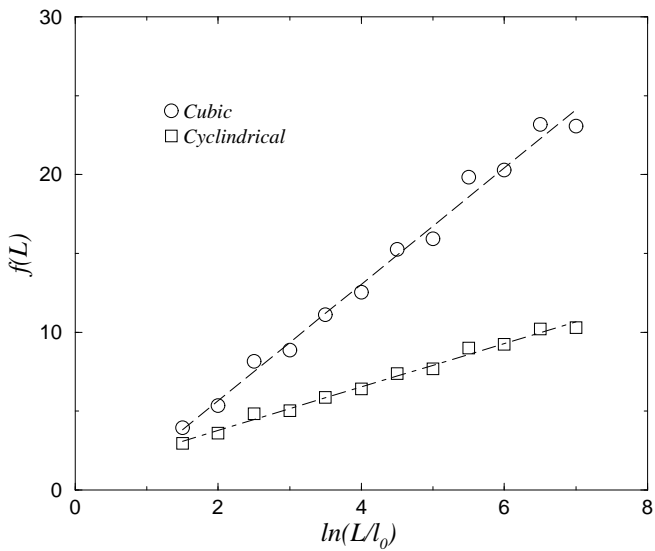}
\caption{The function $f(L)$ in $D(L)\simeq f(L)/L$ at $k=k_c$ is plotted versus $\ln (L/\ell_0)$ for a cubic of length $L$ and  a
cylindrical slab of thickness $L$ and radius $R=10L$.}
\end{figure}

Here we clearly see that the scaling behavior of an intrinsic transport parameter in slabs can be very different from that in spherical
systems due to different sets of recurrent scattering paths. Eqs. (13) and (17) indicate that the reduction of $D(L)/D_0$ at $k=k_c$ due to
WL in slabs is smaller than that in spherical systems by a factor of $\ln L$. This can be explained by the difference between Eq. (6) and
Eq. (16) in volume of integration of the diffusion pole in momentum space. Fig. 1a shows the schematic diagram for the momentum space of the
cylindrical system. The rectangles shown here are the surfaces for volume of evolution about the $q_{\perp}$ axis. The shaded region
represents the momentum space that allows diffusion, which is schematically equal to the total volume of the whole largest square minus the
two grey rectangles and the small white rectangle. This can also be seen from Eqs. (7) and (8). The term of $\eta_A$ represents the largest
square with a length equal to the upper cutoff of $\ell_0^{-1}$ in Eq. (6), which determines the mobility edge, i.e. the value of $k_c$.
$|\eta_B|+|\eta_C|-|\eta_D|$ represents the sum of the grey and white rectangles, which are to be excluded from diffusion. For 'cubic-like'
samples or slabs, the value of $\eta_B$ dominates in Eq. (7) and this is the term which makes these samples different from the spherical
samples. The exclusion of the grey regions implies that diffusion is not allowed when the momentum in each direction is smaller than the
lower momentum cutoffs, i.e. $q_{\parallel}<1/R$ and $q_{\perp}<1/L$. This restriction effectively decreases the total return probability
and thus reduces the effect of WL significantly. The same argument also applies to the cubic geometry, which also involves the separation of
variables in momentum space into two or more directions.

However, there is no such restriction in spherical system. The schematic diagram for momentum space of the spherical system is shown in Fig.
1b. The circles shown are the cross-sections of the concentric spheres about the origin. The shaded area shows the region of allowed
diffusion modes, which is represented by the volume integral in Eq. (16). In spherical geometry, the isotropy of the momentum space retains
the most recurrent scattering paths and thus has the largest WL effect among all geometries in three dimensions. At $k=k_c$, the static
diffusion constant depends only on $L$ and it can be written as $D(L)\simeq f(L)/L$. Eq. (17) indicates $f(L)$ is constant in spherical
system while Eq. (13) indicates $f(L)\propto \ln L$ in cylindrical system. Although we do not have a simple analytical expression of $D(L)$
for cubic system, we also expect $f(L)\propto \ln L$ in a cube. In Fig. 2, we plot the function $f(L)$ versus $\ln (L/\ell_0)$ for a
cylindrical slab of thickness $L$ with radius $R=10L$ and a cube of length $L$. In this graph we can see that $f(L)\propto \ln L$ and $D(L)$
is indeed proportional to $\ln L/L$ in both cylindrical and cubic systems as expected.

\end{subsection}

\begin{subsection}{B. The scaling behavior of wave propagation through disordered slabs at mobility edge}
To study the scaling behavior in dynamics of wave propagation, we consider a pulsed plane wave normally incident on the front surface of a
disordered slab of thickness $L$ at $z=0$. We assume that there is no gain or absorption in the medium and that scattering is isotropic. The
physical quantity we are interested here is the ensemble averaged intensity $\langle I(t,\mathbf{r})\rangle$, which can be obtained from the
Fourier transform of the field-field correlation function in frequency, i.e.,
$C_{\Omega}(\omega;\mathbf{r})=\langle\phi_{\Omega^{+}}(\mathbf{r}) \, \phi^{*}_{\Omega^{-}}(\mathbf{r})\rangle$ and \cite{Drake89,Zhang99}
\begin{equation}
\langle I(t,\mathbf{r})\rangle=\frac{1}{2\pi}\int d\omega \exp(-i \omega t) C_{\Omega}(\omega;\mathbf{r}),
\end{equation}
where $\Omega^{\pm}=\Omega\pm\omega/2$, $\Omega$ is the central frequency, $\omega$  is the modulation frequency and
$\phi_{\Omega}(\mathbf{r})$ is the wave field at position $\mathbf{r}$ inside the sample with a frequency $\Omega$.
$C_{\Omega}(\omega;\mathbf{r})$ can be obtained by solving for the space-frequency correlation function in the following B-S equation,
%%%%%%%%%%%%%%%%%%%%%%%%%%%%%%%%%%%%%%%%%
\begin{eqnarray}
C_{\Omega}(\omega;\mathbf{r},\mathbf{r}')\!\!\! &=&\!\!\!\langle\phi_{\Omega^{+}}(\mathbf{r})\rangle \,
\langle\phi^{*}_{\Omega^{-}}(\mathbf{r}')\rangle \nonumber \\ + \int\!\!\!\! &d\mathbf{r}_{1}&\!\!\! d\mathbf{r}_{2} \, d\mathbf{r}_{3} \,
d\mathbf{r}_{4} \langle G_{\Omega^{+}} (\mathbf{r},\mathbf{r}_{1})\rangle\langle \,
G^*_{\Omega^{-}}(\mathbf{r}',\mathbf{r}_{3})\rangle \, \nonumber \\
&\times& \!\!\! U_{\Omega}(\omega \, ;\mathbf{r}_{1},\mathbf{r}_{2} \, ;\mathbf{r}_{3},\mathbf{r}_{4}) \, C_{\Omega}(\omega
;\mathbf{r}_{2},\mathbf{r}_{4}) \, , \label{} \end{eqnarray}
%%%%%%%%%%%%%%%%%%%%%%%%%%%%%%%%%%%%%%%%
where $\langle\phi_{\Omega}(\mathbf{r})\rangle$ is the coherent source inside the sample, and $\langle G_{\Omega}(\mathbf{r},\mathbf{r}_{1})
\rangle = -{\exp(i\kappa|\mathbf{r}-\mathbf{r}_{1}|)\over4\pi|\mathbf{r}-\mathbf{r}_{1}|}$ is the ensemble-averaged Green's function that
represents the coherent part of wave propagation from $\mathbf{r_1}$ to $\mathbf{r}$ \cite{Lagendijk88}. The complex wavevector
$\kappa=k+{i\over 2\ell}$ describes the ballistic propagation inside the disordered slab, where $k={\Omega\over v}$ is the wavevector, $v$
is the phase velocity, and $\ell$ is the scattering mean free path, which is determined from the imaginary part of the self-energy of
$\langle G\rangle$. In the absence of WL, the bare mean free path, $\ell_0$ , is determined from the single-scattering diagram via
$\ell_0=1/n\sigma$, where $n$ is the density of scatterers and $\sigma$ is the total scattering cross section. The vertex function
$U_{\Omega}$ represents the sum of all irreducible vertices. Here we approximate $U_{\Omega}$ as
%%%%%%%%%%%%%%%%%%%%%%%%%%%%%%%%%%%%%%%%%
\begin{eqnarray}
\lefteqn{U_{\Omega}(\omega;\mathbf{r}_{1},\mathbf{r}_{2};
\mathbf{r}_{3},\mathbf{r}_{4})=}\nonumber\\
& &{4\pi\over\ell_{0}} [1+\delta_L(\omega,k)]\, \delta(\mathbf{r}_{1}-\mathbf{r}_{2}) \, \delta(\mathbf{r}_{1}-\mathbf{r}_{3}) \,
\delta(\mathbf{r}_{1}-\mathbf{r}_{4}). \label{}
\end{eqnarray}
The first term in the vertex function with a scattering strength $4\pi/\ell_{0}$ represents self-avoiding paths and generates all the ladder
diagrams that give rise to wave diffusion when $L\gg \ell_0$ \cite{Lagendijk88}. The second term with a vertex strength
$4\pi\delta_L(\omega,k)/\ell_0$ represents WL contribution to the vertex function \cite{Zhang04}. The presence of this term renormalizes the
bare mean free path to a frequency-dependent mean free path, i.e., $\ell_L(\omega,k)=\ell_{0}/[1+\delta_L(\omega,k)]$. For flux conservation
to hold, the Ward Identity \cite{Vollhardt80,Kirkpatrick85} requires that the mean free path $\ell$ that appears in $\langle G\rangle$
should also be replaced by the same $\ell_L(\omega,k)$.  Since the second term represents recurrent scatterings, it is obtained by summing
all maximally-crossed diagrams due to weak localization. The renormalization factor $\delta_L(\omega,k)$ is obtained by solving Eq. (2) with
appropriate boundary conditions and for the slab geometry considered here, we can use Eq. (5) by taking the limit of $R\rightarrow\infty$.

By using a pulsed plane-wave excitation, the averaged intensity is uniform over the transverse cross-section of the slab and Eq. (20) can be
expressed as \cite{Zhang99}
\begin{eqnarray}
C_{\Omega}(\omega\!\!\!\!\!&,&\!\!\!\!z)=
\exp\!\left({i\omega z\over v}-{z\over\ell_L(0)}\right) \nonumber \\
&+&\!\!\frac{1}{4\pi\ell_L(\omega,k)}\int_{0}^{L}\!\!dz' H(\omega,z-z')\, C_{\Omega}(\omega,z') \, ,
\end{eqnarray}
where $\ell_L(0)\equiv\ell_L(0,k)$ and
\begin{eqnarray}
H(\omega ,z-z')&=&\pi \int d\rho^{2} \nonumber \\
&\times&{\exp \!\left[ \left({i\omega\over v} -{1\over\ell_L(\omega,k)}\right)\!\sqrt{\rho^{2}+(z-z')^{2}}\right] \over
\rho^{2}+(z-z')^{2}}.\nonumber \\ \,
\end{eqnarray}
Eq. (22) is then numerically solved for $C_{\Omega}(\omega,z)$.  The transmitted intensity $\langle I(t,L)\rangle$ is calculated from the
Fourier transform of $C_{\Omega}(\omega,L)$ as in Eq. (19) and $C_\Omega(0,L)$ gives the static transmitted intensity $\langle I(L)\rangle$.

In the second method to obtain $\langle I(t,\mathbf{r})\rangle$, we solve the diffusion equation in the frequency domain with a frequency
dependent diffusion constant, $D_L(\omega,k)$. The WL effects are incorporated through $D_L(\omega,k)$ according to Eqs. (4) and (5) with
$R\rightarrow\infty$. The solution to the diffusion equation in a slab takes the form

\begin{equation}
C_{\Omega}(\omega,z)=\frac{2}{\tilde{L}}\sum_{n=1}^{\infty}\frac{\sin[q_n(\overline{z}_e+\overline{z}_p)]\sin[q_n(z+\overline{z}_e)]}{-i\omega+D_L(\omega,k)q_n^2},
\end{equation}
where $q_n=n\pi/\tilde{L}$ is the transverse momentum, $\tilde{L}=L+2\overline{z}_e$ is the effective length,
$\overline{z}_e\simeq0.71\ell_L(0,k)$ is the extrapolation length and $\overline{z}_p\simeq\ell_L(0,k)$ is the penetration length. Here we
use the renormalized mean free path $\ell_L(0,k)=\ell_0/(1+\delta_L(0,k))$ in the evaluation of the extrapolation length $\overline{z}_e$
and the penetration length $\overline{z}_p$. This replacement is consistent with the replacement of $\ell_0$ by $\ell_L(\omega,k)$ in both
the averaged green function $\langle G\rangle$ and vertex function $U_\Omega$ in the B-S equation as required by the Ward identity.  It
should be mentioned that the dynamic diffusion constant $D(\omega)$ has been studied for electrons near the mobility edge
\cite{Gotze,Shapiro}. In these studies, a behavior of $D(\omega)\propto \omega^{1/3}$ was found at mobility edge for an unbounded medium.

The static transmitted intensity $\langle I(L)\rangle$ can be obtained from Eq. (24) by setting $\omega=0$, yielding
\begin{equation}
\langle I(L)\rangle=\frac{\overline{z}_e+\overline{z}_p}{L+2\overline{z}_e}\frac{\overline{z}_e}{D_L(k)}.
\end{equation}
The transmission $\langle T(L)\rangle=-D_L(k)\frac{d}{dz}\langle I(z)\rangle|_{z=L}$ and has the form

\begin{equation}
\langle T(L)\rangle=\frac{\overline{z}_e+\overline{z}_p}{L+2\overline{z}_e}.
\end{equation}
At $k=k_c$, Eqs. (25) and (26) give the same scaling behavior of $\ln L/L^2$ for both $\langle I(L)\rangle$ and $\langle T(L)\rangle$. It is
worth to note that it is the replacements of $\overline{z}_p\simeq\ell_0$ by $\overline{z}_p\simeq\ell_L(0,k)$ and
$\overline{z}_e\simeq0.71\ell_0$ by $\overline{z}_e\simeq0.71\ell_L(0,k)$ that change the scaling of $\langle I(L) \rangle$ and $\langle
T(L)\rangle$ from $1/L^2$ to $\ln L/L^2$.

\end{subsection}
\end{section}

\begin{subsection}{C. Discussion on the position-dependent diffusion constant}
In an open system, the WL effects should vary in space as the probability of returning to each point inside the sample can be different.
Thus in Eq. (1) the renormalized diffusion constant $D$ can also be position dependent. In arriving Eq. (12) for slab geometry, we have
simplified the calculation by taking the spatial average of $\tilde{G}$ along the $z$-axis and assumed that the diffusion constant is
independent of $z$. Effectively, this simplification replaces $D(z)$ by its harmonic mean, i.e., $\overline{D}(L)\equiv
\langle1/D(z)\rangle^{-1}$ as can be seen from Eq. (1). A complete theory requires self-consistent solutions of both $D(z)$ and
$\tilde{G}(\mathbf{r},\mathbf{r'})$, from which one can obtain $\langle T(L)\rangle$. In the case of slab geometry and static limit, Eq. (2)
is replaced by the following position-dependent diffusion equation \cite{Bart00}, i.e.,

\begin{figure}
\includegraphics [width=\columnwidth] {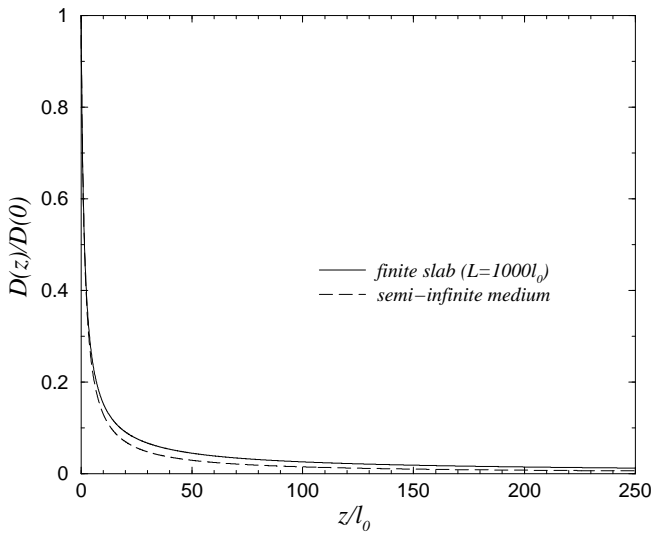}
\caption{$D(z)/D(0)$ for a finite slab with $L=1000\ell_0$ (solid line) and the semi-infinite medium (dashed line), which are obtained from
Eqs. (32) and (31) respectively, are plotted as functions of $z/\ell_0$.}
\end{figure}

\begin{equation}
\frac{d}{dz}\left[D(z)\frac{d
\tilde{G}(\mathbf{q_{\parallel}};z,z')}{dz}\right]-D(z)q^2_{\parallel}\tilde{G}(\mathbf{q_{\parallel}};z,z')=-\delta(z-z'),
\end{equation}
where
\begin{equation}
\tilde{G}(\mathbf{r},\mathbf{r}')=\frac{1}{(2\pi)^2}\int d\mathbf{q_{\parallel}}
\exp(i\mathbf{q_{\parallel}}\cdot\mathbf{\rho})\tilde{G}(\mathbf{q_{\parallel}};z,z').
\end{equation}
In an operator form, we can write Eq. (27) as $\mathcal{L}\{\tilde{G}(\mathbf{q_{\parallel}};z,z')\}=-\delta(z-z')$, where the self-adjoint
linear operator $\mathcal{L}$ is given by

\begin{equation}
\mathcal{L}\equiv \frac{d}{dz}\left[D(z)\frac{d}{dz}\right]-D(z)q^2_{\parallel}.
\end{equation}
For a rigorous approach, one should solve Eq. (1) and Eq. (27) simultaneously with appropriate boundary conditions at two surfaces for each
fixed sample thickness $L$. The transmission coefficient is then obtained from the relation \cite{Bart00}
\begin{equation}
\langle T(L)\rangle=(z_e+z_p)\left[2z_e+\int_0^L dz\frac{D_0}{D(z)}\right]^{-1},
\end{equation}
where the extrapolation length $z_e$ and the penetration length $z_p$ are determined by the diffusion constant at sample boundary
\cite{Bart00}. By repeating the same calculation at different $L$'s, one can obtain the scaling behavior of $\langle T(L)\rangle$. For
$L\gg\ell_0$, Eq. (30) can be written as  $\langle T(L)\rangle\simeq (z_e+z_p)\left[\int_0^L dz\frac{D_0}{D(z)}\right]^{-1}\simeq
\frac{z_e+z_p}{D_0}\frac{\overline{D}(L)}{L}$. By comparing Eq. (30) with Eq. (26), we can see that the scaling of $\langle T(L)\rangle$ for
both position-dependent and position-independent diffusion constant are determined by the scaling of $\overline{D}(L)$. However, there is a
subtle difference between the two situations. In Eq. (26) the scaling of $\langle T(L)\rangle$ is affected by the scaling of
$\overline{z}_e$ and $\overline{z}_p$ through the scaling of $\ell_L$ or $\overline{D}(L)$. However, in Eq. (30) the scaling of $\langle
T(L)\rangle$ arises directly from $\overline{D}(L)/L$. Since $\langle T(L)\rangle$ is dominated by the small values of $D(z)$ deep inside
the sample, an accurate numerical calculation for $\langle T(L) \rangle$ is difficult when $L$ is large. In this work, we do not intend to
solve this problem self-consistently. Instead we would like to propose a plausible form of $D(z)$ and show that the corresponding
transmission is close to the self-consistent solution and behaves like $\langle T(L)\rangle\propto \ln L/L^2$ when $L$ is large.

For a semi-infinite medium, van Tiggelen et. al. have performed the self-consistent calculations discussed above and suggest an analytical
form of $D(z)$ at mobility edge \cite{Bart00}, i.e.,
\begin{equation}
D_\infty(z)=\frac{D_\infty(0)}{1+z/\xi_c},
\end{equation}
where $D_\infty(0)$ is the diffusion constant at the boundary of a semi-infinite medium. Eq. (31) shows $D_\infty(z)$ decreases like $1/z$
from its value at the boundary when $z$ is moving into the semi-infinite medium. In the absence of internal reflection, they find
$D_\infty(0)/D_0=0.642$ and $\xi_c/\ell_0=1.5$. For a finite slab of thickness $L$, based on Eq. (31), they have also suggested that
$D_L(z)\simeq D_\infty(z_L)$, where $z_L=\frac{L}{2}-|\frac{L}{2}-z|$ \cite{Bart00}, from which they found $\langle
T(L)\rangle\propto1/L^2$.  In the case of finite slabs, we expect that $D_L(z)$ should decrease slower than $1/z$ as $z$ moves well inside
the sample due to reduced WL effects in the presence of other boundary.  By taking this into account, we propose here the following modified
form for $D_L(z)$:

\begin{equation}
D_L(z)=D_\infty(0)\left\{\left[1+\frac{(z_L/\ell_0)(2\ln(z_L/\ell_0+w)-1)}{(\ln(z_L/\ell_0+w))^2}\right]^{-1}\right.
\end{equation}
with $L\gg w\gg\xi_c$. The above $D_L(z)$ is symmetric with respect to the central plane at $L/2$. Its value decreases monotonically from
the boundary to its minimum at the center. In this work, we choose $w\simeq10.65$ in Eq. (32) to match the behavior of Eq. (31) near the
sample boundaries with $\xi_c/\ell_0=1.5$. In order to make comparison between Eqs. (31) and (32), In Fig. 3 we plot both
$D_L(z)/D_\infty(0)$ of Eq. (32) for the case of a finite slab with $L/\ell_0=1000$ (solid line) and $D_\infty(z)/D_\infty(0)$ of Eq. (31)
for semi-infinite medium (dashed line). From Fig. 3, it is easy to see $D_L(z)$ decays like $D_\infty(z)$ near the boundary, but decreases
in a slower rate than $1/z$ away from the boundary. Since the value of $\langle T(L)\rangle$ is dominated by the small values of $D_L(z)$
inside the sample as it can be seen from Eq. (30), a scaling behavior which is different from $\langle T(L)\rangle\propto1/L^2$ is expected.
By substituting Eq. (32) into Eq. (30) and setting $z_p=3D_\infty(0)/v=0.642\ell_0$ and $z_e=0.71z_p$ \cite{Bart00}, we obtain the
transmission, $\langle T^0(L)\rangle$, which is shown by the dashed curve in Fig. 4.  In order to show more clearly its scaling behavior, we
replot the function $(\langle T^0(L)\rangle)^{-1}\ln L$ in the inset of Fig. 4 using log-log scale.  A linear line of slope 1.98 clearly
shows the scaling relation $\langle T(L)\rangle\propto \ln L/L^2$.

\begin{figure}
\includegraphics [width=\columnwidth] {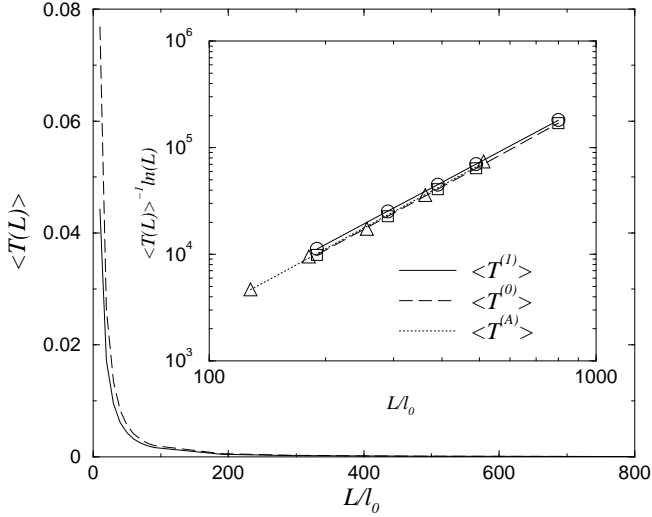}
\caption{$\langle T^{(1)}(L)\rangle$ obtained from the first iteration of Eq. (33) (solid line) and $\langle T^{(0)}(L)\rangle$ obtained
directly from $D_L(z)$ in Eq. (32) (dashed line) are plotted against $L/\ell_0$. In the inset, $\langle T^{(1)}\rangle^{-1}\cdot(\ln L)$
(solid line) and $\langle T^{(0)}\rangle^{-1}\cdot(\ln L)$ (dashed line) are plotted versus $L/\ell_0$ in log-log scale. The slope of their
linear fitting lines are 1.95 and 1.98 respectively. Also plotted in the inset is $\langle T^{(A)}\rangle^{-1}\cdot(\ln L)$ obtained from
Eq. (26) (dotted line) with a linear fitting line of slope 1.97.}
\end{figure}

\begin{figure}
\includegraphics [width=\columnwidth] {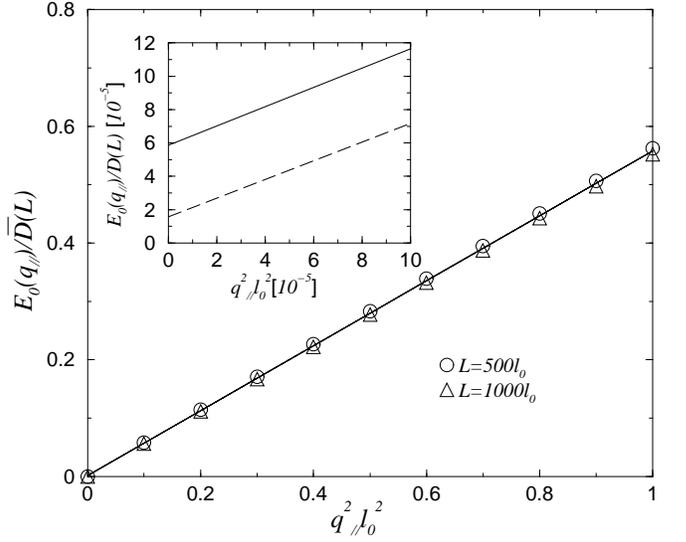}
\caption{The quantity $E_0(\mathbf{q_{\parallel}})/\overline{D}(L)$ for $L/\ell_0=500$ and $1000$ are plotted against
$q^2_{\parallel}\ell_0^2$.  In the inset, the curves near $q^2_{\parallel}\ell_0^2\simeq0$ for $L/\ell_0=500$ (solid line) and $1000$
(dashed line) are amplified by a factor of $10^5$ to show the y-intercepts. The fit to the curves suggest that
$E_0(\mathbf{q_{\parallel}})\simeq E_0(0)+\overline{D}(L)\beta q_{\parallel}^2$ with $\beta\simeq0.55$.}
\end{figure}

\begin{figure}
\includegraphics [width=\columnwidth] {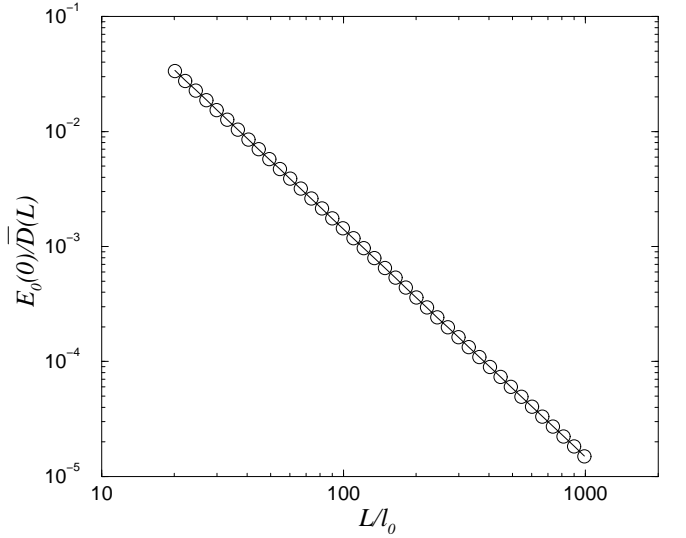}
\caption{The values of $E_0(0)/\overline{D}(L)$ are plotted against $L/\ell_0$ in log-log scale. The slope and the y-intercept of the linear
fitting  are -1.99 and 2.59 respectively, thus $E_0(0)\simeq1.35\overline{D}(L)(\pi/L)^2$.}
\end{figure}

In order to test the self-consistency of this result, we substitute Eq. (32) into Eq. (29) and solve for the eigenvalue problem of
$\mathcal{L}\{U_n(\mathbf{q_{\parallel}},z)\}=E_n(\mathbf{q_{\parallel}})\{U_n(\mathbf{q_{\parallel}},z)\}$ by using the method described in
Ref. \cite{Brad98}. The diagonal Green's function $\tilde{G}(\mathbf{r},\mathbf{r})$  of Eq. (27) can be obtained by the eigenfunction
expansion:

\begin{equation}
\tilde{G}(z,z)=\frac{1}{(2\pi)^2}\int d\mathbf{q_\parallel}\sum_{n=0}^{n_c}
\frac{|U_n(\mathbf{q_{\parallel}},z)|^2}{E_n(\mathbf{q_{\parallel}})}.
\end{equation}
In Sec. IIA, we have shown that the $\ln L$ factor found in $\eta_B$ of Eq. (8) arises from the exclusion of phase space volume bounded by
the lowest order mode $q_\perp=\pi/L$ when $\mathbf{q_\parallel}\simeq 0$ as shown in the grey area of Fig. 1. Here we would like to show
that this volume exclusion effect holds even when a position-dependent $D(z)$ is considered. Since we are interested in the scaling of
$\langle T(L)\rangle$, only the spatial average of $\tilde{G}(z,z)$ is relevant as can be seen from Eq. (30). The normalization of
eigenfunctions requires $\overline{|U_n(\mathbf{q_{\parallel}})|^2}=1/L$ for all $n$ and $\mathbf{q_\parallel}$. We can see that the
forbidden volume in phase space is determined by $E_n(\mathbf{q_{\parallel}})$. For a fixed $\mathbf{q_\parallel}$, the lowest order
eigenvalues for different slab thickness $L$ divided by $\overline{D}(L)$, i.e. $E_0(\mathbf{q_{\parallel}})/\overline{D}(L)$, are plotted
versus $q^2_{\parallel}\ell_0^2$ in Fig. 5. The y-intercepts for each curves are also shown in the inset of Fig. 5. The linear fits to the
curves for each $L$ suggest that $E_0(\mathbf{q_{\parallel}})\simeq E_0(0)+\overline{D}(L)\beta q_{\parallel}^2$. For sufficiently large
$L$, $\beta\simeq0.55$. By using the y-intercepts of the fitted lines in Fig. 5, $E_0(0)/\overline{D}(L)$ are also plotted against
$L/\ell_0$ in log-log scale in Fig 6, which are well fitted by the formula $E_0(0)\simeq1.35\overline{D}(L)(\pi/L)^2$. The combined results
indicates that $E_0(\mathbf{q_{\parallel}})\simeq \overline{D}(L)[1.35(\pi/L)^2+\beta q_{\parallel}^2]$ and the exclusion of phase space
volume is retained for finite $L$. By using this result in Eq. (33), we would expect to obtain the same scaling behavior for the
transmission. In order to confirm our assertion, we first calculate the averaged diffusion constant, $\overline{D}^{(1)}(L)$ from Eq. (1) by
using the results shown in Figs. 5 and 6. By substituting $\overline{D}^{(1)}(L)$ into Eq. (30) we obtain $\langle T^{(1)}(L)\rangle$. This
result is plotted as the solid curve in Fig. 4. The excellent agreement between $\langle T^{(0)}(L)\rangle$ and $\langle T^{(1)}(L)\rangle$
when $L>100\ell_0$ indicates that Eq. (32) is close to the self-consistent solution when $L/\ell_0$ is large. This result strongly indicates
that the scaling behavior of $D(L)\propto \ln L/L$ or $\langle T(L)\rangle\propto \ln L/L^2$ found in the previous section holds even when a
position-dependent diffusion constant is considered.  In fact, we will show in Sec. III that the transmission coefficient shown in Fig. 4
agrees very well with that obtained from an averaged diffusion constant $D(L)$ given in Eq. (14) through the use of Eq. (26), which is shown
by the dotted line in the inset of Fig. 4. In the next section, we present the numerical results of wave propagation through disordered
slabs at mobility edge based on the averaged frequency-dependent diffusion constant $D_{L}(\omega,k)$ shown in Eqs. (4) and (5).
\end{subsection}

\begin{section}{III. Numerical results and discussions}
Before presenting the dynamic results, we would like to show that Eq. (25) of the diffusion approximation (DA) is capable of producing the
results of the B-S equation when $L\gg\ell_0$. We first consider a case in the diffusive regime with $k>k_c$. In Fig. 7, we plot the
reciprocal of static transmitted intensity $\langle I(L)\rangle^{-1}$ at $k\ell_0=8$ against the dimensionless slab thickness $L/\ell_0$ in
log-log scale. Both the results from the B-S equation and  the DA are shown. It can be seen that the results from the DA agrees well with
those from the B-S equation apart from a small constant shift. The dashed line with a slope of $1$ is plotted to show that $\langle
I(L)\rangle\propto1/L$ for both the B-S and the DA results. At the mobility edge, i.e. $k=k_c$, we would expect $\langle I(L)\rangle\propto
\ln L/L^2$.  On the right scale of Fig. 7, we plot the calculated result of $\langle I(L)\rangle^{-1}\cdot(\ln L)$. The solid line with a
slope of $2$ is also plotted to show that $\langle I(L)\rangle$ calculated from both the B-S equation and the DA indeed give the scaling
behavior of $\ln L/L^2$. We have also calculated the transmission coefficient using Eq. (26).  These results are denoted as $\langle
T^{(A)}\rangle$ and plotted as the dotted line in the inset of Fig. 4. The excellent agreement between $\langle T^{(A)}\rangle$ and $\langle
T^{(0)}\rangle$ or $\langle T^{(1)}\rangle$ supports the use of an averaged diffusion constant in the transmission calculations.

\begin{figure}
\includegraphics [width=\columnwidth] {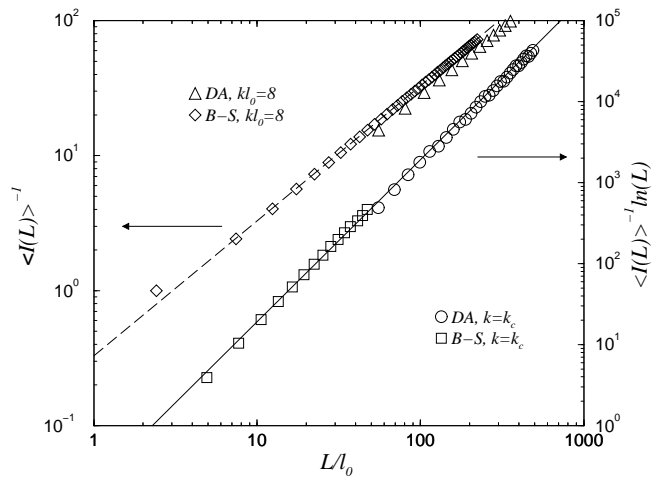}
\caption{The reciprocal of static transmitted intensity $\langle I(L)\rangle^{-1}$ for $k\ell_0=8$ is plotted as a function of $L/\ell_0$ in
log-log scale. Both B-S results and diffusion approximation (DA) results are shown. The dashed line with slope of $1$ is plotted to shown
$\langle I(L)\rangle\propto 1/L$. The scaled reciprocal static transmitted intensity $\langle I(L)\rangle ^{-1}\cdot(\ln L)$ for $k=k_c$ is
also plotted with the logarithmic scale on the right. The solid line with slope of $2$ shows that $\langle I(L)\rangle\propto\ln L/L^2$ at
$k=k_c$.}
\end{figure}

\begin{figure}
\includegraphics [width=\columnwidth] {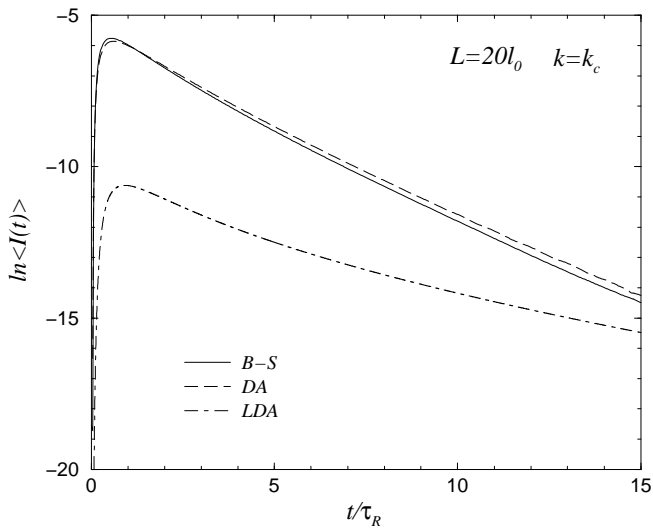}
\caption{The logarithmic time-resolved transmitted intensities $\ln\langle I(t) \rangle$ at $k=k_c$ for $L/\ell_0=20$ obtained from the B-S
equation (in solid line) and diffusion approximation (in dashed line) are plotted as functions of $t/\tau_R$. The numerical result from the
local diffusion approximation (in dot-dashed line) with the same parameters is also plotted for comparison.}
\end{figure}

In Fig. 8, we plot $\langle I(t)\rangle\equiv \langle I(t,L)\rangle$ for $L/\ell_0=20$ at $k=k_c$ obtained from the B-S equation (in solid
line) and the DA (in dashed line) as functions of time $t/\tau_R$, where $\tau_R=\tilde{L}^2/\pi^2D(L)$ is the renormalized diffusion time.
Our results show that they agree with each other for a rather large range of $t/\tau_R$, indicating the validity of Eq. (24) when $L\gg
\ell_0$. In the same graph, we also plot the result obtained from Eq. (12) of Ref. \cite{Berkovits90}, in which a time-dependent diffusion
constant $D(t)=D_0(\ell_0/tv)^{1/3}$ is used in the time-dependent diffusion equation. Since such a local scaling approach does not consider
the retardation effect of the recurrent scattering paths, it overestimates the WL effects and, therefore, produces a smaller $\langle
I(t)\rangle$ and with a slower decay rate as shown in Fig. 8. Such approach has also been used in the study of coherent backscattering
\cite{Berkovits} and absorbing media \cite{Yosefin} near mobility edge. In our theory, it is the frequency-dependence of the factor
$\delta_L$ in Eq. (5) that makes the reduction of intrinsic diffusion constant causal in time. It is also interesting to point out that,
unlike a pure diffusion process, the decay of $\langle I(t)\rangle$ shown in Fig. 8 is not a simple exponential decay.  The slowdown of
decay rate in time is a result of increasing WL effect contributed by the presence of longer recurrent scattering paths. Such
non-exponential decay has also been reported first for electronic systems \cite{Altshuler, Mirlin00} and recently observed in the microwave
experiments in a nominally diffusive region \cite{Chabanov02}. Here the cause of the non-exponential decay is also due to WL effects, but in
a quasi-1D geometry \cite{Zhang04, Mirlin00, Bart04}

\begin{figure}
\includegraphics [width=\columnwidth] {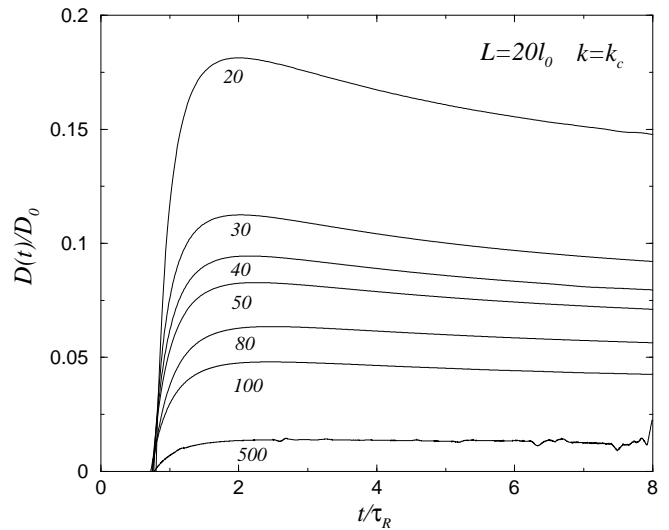}
\caption{The time-dependent diffusion coefficient $D(t)/D_0$ at $k=k_c$ for $L/\ell_0=20$, $30$, $40$, $50$, $80$, $100$ and $500$, obtained
from the diffusion approximation, are plotted as functions of time $t/\tau_R$.}
\end{figure}

The non-exponential decay shown in Fig. 8 becomes more transparent in the time-dependent diffusion coefficient $D(t)$, which is defined via
the relation $D(t)/D_0=-\tau_D\,d\ln\langle I(t,L)\rangle/dt$, for $t>\tau_R$, where $\tau_D\equiv \tau_R/(1+\delta_L)$ is the diffusion
time \cite{Chabanov02}. In Fig. 9 we plot $D(t)/D_0$ at $k=k_c$, obtained from the DA, for various sample thicknesses $L/\ell_0$ as
functions of $t/\tau_R$. It is easy to see that $D(t)/D_0$ is not a constant in time and its maximum value decreases as $L$ increases. This
is in contrast to the case of the diffusive regime, i.e. when $k>k_c$, in which the WL effect is weak and the change of slope in $D(t)$ is
very small \cite{Zhang_C04}.

Since the static diffusion constant $D(L)$ has a scaling that is different from that of the spherical samples, it is also interesting to
investigate the scaling of the time-dependent diffusion coefficient $D(t)$. Here we are interested in $D(t)$ in the long time limit, i.e.
$D(t\rightarrow\infty)$, because the WL contribution of the long recurrent scattering paths is expected to saturate eventually. In the inset
of Fig. 10. we show the fitting of the long time tail of $D(t)/D_0$ (in solid line) against the function $a+b/t$ in dashed line for the case
of $L=20\ell_0$ at $k=k_c$. We perform the similar fitting to each of the curves in Fig. 9. From these fittings, we obtain
$D(t\rightarrow\infty)=a$ for different slab thickness $L/\ell_0$ and they are used to plot $D(t\rightarrow\infty)/D_0\ln L$ as a function
of $L/\ell_0$ in log-log scale in Fig. 10. The solid line is the fit to the curve with a slope of $-0.99$, which suggests that
$D(t\rightarrow\infty)/D_0\propto\ln L/L$ and is consistent with the scaling of $D(L)$ given by Eq. (13).

Figs. 9 and 10 show that the $\ln L$ factor appears when waves see the boundary of the sample, i.e., $t\gtrsim\tau_R$. When $t<\tau_R$,
waves have not reached the output surface and the $\ln L$ factor should not appear. In fact, we find $D(\omega)$ behaves like $\omega^{1/3}$
when $\omega > 1/\tau_R$. This is consistent with the previous studies \cite{Gotze,Shapiro}. When $\omega<1/\tau_R$, we find $D(\omega)
-D(L) \propto \omega^2$, which represents the saturation of WL effects, or equivalently, the saturation of $D(t)$ when $t/t_R \gg 1$ as
shown in the inset of Fig. 10.

\begin{figure}
\includegraphics [width=\columnwidth] {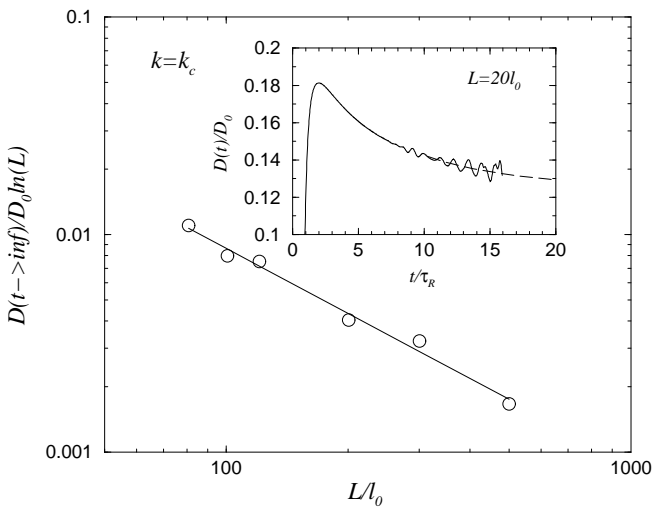}
\caption{The scaled long time diffusion coefficients $D(t\rightarrow\infty)/D_0\ln L$ at $k=k_c$, obtained from the fitted values of $a$
from the curves in Fig. 5, are plotted as a function of $L/\ell_0$ in log-log scale. The slope of the fitted line is -0.99. In the inset,
the dashed line shows the fitting of $D(t\rightarrow\infty)/D_0$ (in solid line) by the function $a+b/t$ for the case of $L=20\ell_0$.}
\end{figure}

\begin{figure}
\includegraphics [width=\columnwidth] {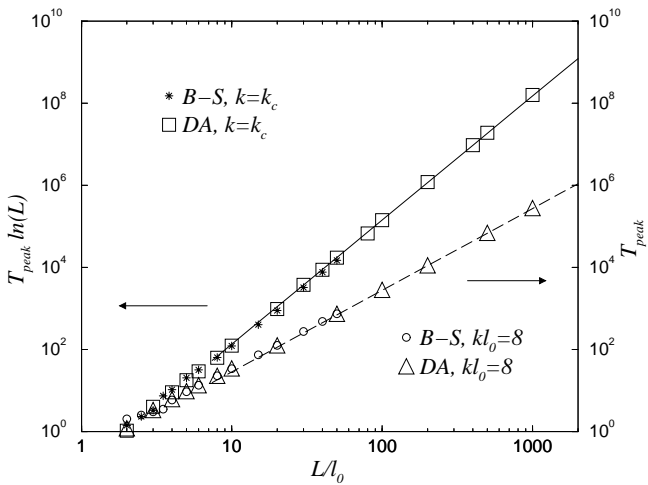}
\caption{The time of peak intensity $T_{peak}(L)$ at $k\ell_0=8$ is plotted against $L/\ell_0$ in log-log scale, with the logarithmic scale
shown on the right. Both B-S results and diffusion approximation (DA) results are shown. The slope of the fitted line (dashed line) is
$1.99$. The scaled time of peak intensity, $T_{peak}\ln L$,  at $k=k_c$ versus $L/\ell_0$ is also plotted with the logarithmic scale shown
on the left. The slope of the fitted line (solid line) is $3.01$.}
\end{figure}

\begin{figure}
\includegraphics [width=\columnwidth] {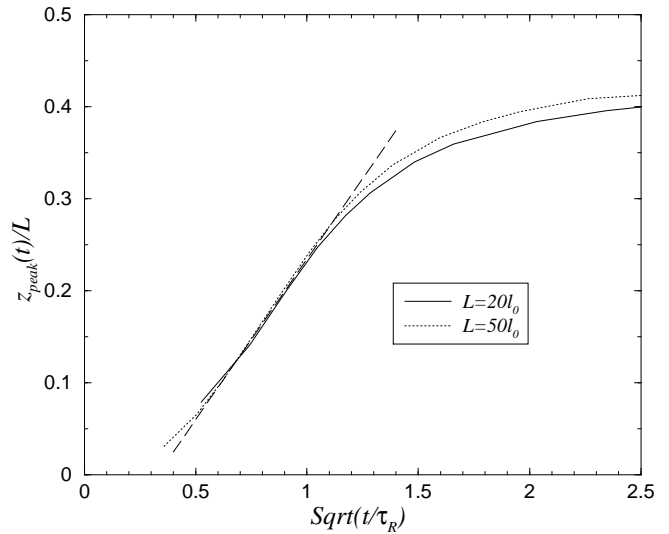}
\caption{The position of peak intensity $z_{peak}(L)/L$ obtained from the B-S equation  is plotted against $\sqrt{t/\tau_R}$. The dashed
line is the fitting to the linear region, which has a slope of $0.33$.}
\end{figure}

Besides the static intensity and the diffusion coefficient, the time of peak intensity $T_{peak}$, i.e. the time it takes waves to diffuse
across the sample, also follows the scaling of $\delta_L(0,k)$. $T_{peak}$ can be obtained from the peak position of the time-resolved
transmitted intensity $\langle I(t)\rangle$. In standard diffusion theory, $T_{peak}\simeq L^2/3D_0$ \cite{Bart99}. If weak localization in
included, the Boltzmann diffusion coefficient $D_0$ should be replaced by $D_L(k)$ and thus $T_{peak}\simeq
(1+\delta_L)L^2/3D_0=(1+\delta_L)L^2/v\ell_0$. $T_{peak}$ versus $L/\ell_0$ at $k\ell_0=8$ in log-log scale is plotted in Fig. 11 with the
logarithmic scale shown on the right. Both B-S and DA results are shown and they agrees with each other except for small sample thickness.
The slope of the dashed line is $1.99$, which confirms that $T_{peak}\propto L^2$ at weak scattering limit. The scaled peak time of
intensity, $T_{peak}\ln L$,  at $k=k_c$ versus $L/\ell_0$ is also plotted on the same graph with the logarithmic scale shown on the left.
The slope of the fitted line is $3.01$, which indicates that $T_{peak}\propto L^3/\ln L$ and is consistent with the scaling of
$\delta_L(0,k_c)\propto L/\ln L$.

We have also studied the position of the peak intensity $z_{peak}$ obtained from B-S equation, which represents the collective transport of
diffusive waves in the sample. If weak localization is present, the recurrent scattering paths enhance the backscattering of waves and thus
delay the propagation of the peak intensity. $z_{peak}(L)/L$ at $k=k_c$ for $L/\ell_0=20$ and $50$ are plotted as functions of
$\sqrt{t/\tau_R}$ in Fig. 12. The fitted line with a slope of $0.33$ is added to highlight the region of diffusive transport. This suggests
that $z_{peak}/L\simeq0.33\sqrt{t/\tau_R}$ and thus $z_{peak}\sim \sqrt{D(L)t}$ for $t<\tau_R$, which is consistent with the diffusion
theory. When $t>\tau_R$, the curves start to deviate from the dashed line because the diffusive waves have reached the open end. As we can
see in Fig. 9, the value of $z_{peak}/L$ is always less than $0.5$ and this implies the wave interference under strong scattering can stop
the peak intensity from approaching the central plane of the slab.
\end{section}

\begin{section}{IV. Conclusions}
In summary, the scaling behavior of wave transport at the Anderson transition, i.e. $k=k_c$ has been studied. We found that both the static
and dynamic transport properties at $k=k_c$ follow the scaling of the averaged static diffusion constant $D(L)$. $D(L)$ is found to scale as
$\ln L/L$ in cubic, cylindrical samples or slabs, in contrast to the scaling of $D(L)\propto 1/L$ found previously for electrons or
spherical samples. The corresponding static transmission $\langle T(L)\rangle$  scales like $\ln L/L^2$, in contrast to the $1/L^2$ behavior
found previously. Our results indicates that the weak localization effects in other geometry are in general weaker than that of the
spherical system by a factor of $1/\ln L$.  This factor arises from the existence of larger volume exclusion in the phase space of allowed
diffusion modes in non-spherical samples.   For dynamic transport, we solved both the Bethe-Salpeter equation and the diffusion equation
with weak localization included self-consistently. The numerical results calculated by the two methods agrees qualitatively with each other
and they are found to produce the same scaling behavior when $L\gg\ell_0$. The scaling for the long time diffusion coefficient and the time
of peak intensity are $D(t\rightarrow\infty)/D_0\propto \ln L/L$ and $T_{peak}\propto L^3/\ln L$ respectively, which are consistent with the
scaling of $D(L)$ in slabs. In addition, the position of peak intensity in a slab $z_{peak}/L$ is found to scale as $\sqrt{t/\tau_R}$ when
$t<\tau_R$, which is also consistent with the diffusion theory of $D(L)$.  We have also studied the position-dependent weak localization
effects by using a plausible form of position-dependent diffusion constant $D(z)$.  The same scaling behavior is obtained for the
transmission, i.e., $\langle T(L)\rangle \propto \ln L/L^2$.
\end{section}

The authors would like to thank B.A. Foreman for the introduction of Ref. \cite{Brad98}. This research is supported by Hong Kong RGC Grant
No.~HKUST 6058/02P.

\end{document}